\documentclass[%
 reprint,
superscriptaddress,
bibnotes,
 amsmath,amssymb,
]{revtex4-2}

\usepackage{graphicx}
\usepackage{dcolumn}
\usepackage{bm}

\begin{document}

\title{Coherent exciton-exciton interactions and exciton dynamics in a MoSe\textsubscript{2}/WSe\textsubscript{2} heterostructure}

\author{Torben L. Purz}
\affiliation{University of Michigan, Department of Physics, Ann Arbor, Michigan 48109, USA}
\author{Eric. W. Martin}
\affiliation{University of Michigan, Department of Physics, Ann Arbor, Michigan 48109, USA}
\affiliation{MONSTR Sense Technologies, LLC, Ann Arbor, Michigan 48104, USA}
\author{Pasqual Rivera}
\affiliation{ Department of Physics, University of Washington, Seattle, Washington 98195-1560, USA}
\author{William G. Holtzmann}
\affiliation{ Department of Physics, University of Washington, Seattle, Washington 98195-1560, USA}
\author{Xiaodong Xu}
\affiliation{ Department of Physics, University of Washington, Seattle, Washington 98195-1560, USA}
\author{Steven T. Cundiff}
 \email{cundiff@umich.edu}
\affiliation{University of Michigan, Department of Physics, Ann Arbor, Michigan 48109, USA}

\date{\today}


\begin{abstract}
    Coherent coupling between excitons is at the heart of many-body interactions with transition metal dichalcogenide (TMD) heterostructures as an emergent platform for the investigation of these interactions.
    We employ multi-dimensional coherent spectroscopy on monolayer MoSe\textsubscript{2}/WSe\textsubscript{2} heterostructures and observe coherent coupling between excitons spatially localized in monolayer MoSe\textsubscript{2} and WSe\textsubscript{2}. Through many-body spectroscopy, we further observe the absorption state arising from free interlayer electron-hole pairs. This observation yields a spectroscopic measurement of the interlayer exciton binding energy of about 250 meV.
\end{abstract}

\maketitle

Two-dimensional semiconducting transition metal dichalcogenides (TMDs) have recently received considerable attention for their efficient light-matter coupling, which has been harnessed in various optoelectronic applications \cite{TMD_Photodiode,TMD_Laser,Optoelectronics}. 
Because of the high tunability of electronic properties, moir\'e superlattice effects for engineering interacting excitonic lattices \cite{Moire1,Moire2}, and the strong Coulomb interactions at the 2D limit \cite{CoherentGalan}, they are an emergent platform for investigating many-body excitonic interactions. The strong Coulomb interactions yield strong coherent and incoherent interactions between different exciton states.
Excitonic coherent coupling is crucial for many-body effects with potential applications such as electromagnetically induced transparency \cite{EMTransparency1,EMTransparency2}, lasing without inversion \cite{LasingInversion1,LasingInversion2,LasingInversion3} and excitonic quantum degenerate gases \cite{ExcitonCondensationReview}.
This coupling has been observed in numerous conventional semiconductors \cite{QW1,QW2,QD1,QD2}.
Moreover, coherent coupling between excitons and trions has recently been observed in MoSe\textsubscript{2} monolayers \cite{CoherentXiadong, CoherentGalan}.

\begin{figure}
\includegraphics[width=0.5\textwidth]{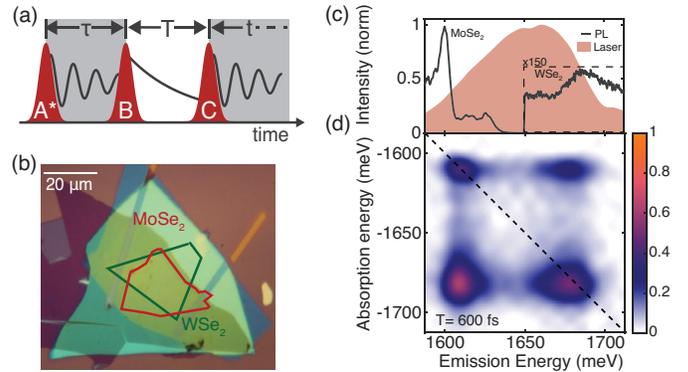}
\caption{\label{fig:Fig1} (a) Schematic of a three-pulse MDCS
experiment (Fourth pulse used for heterodyne detection not shown). (b)  Microscope image of the heterostructure sample (c) Photoluminescence spectrum of the sample, together with the laser excitation spectrum. (d) Characteristic low-temperature, low-power multidimensional coherent spectrum of the MoSe\textsubscript{2}/WSe\textsubscript{2} heterostructure at a pump-probe delay $T$\,=\,600\,fs. The occurence of two off-diagonal coupling peaks suggest the existence of coherent, and/or incoherent coupling channels between the two materials.}
\end{figure}

While for TMD heterostructures incoherent processes, including ultrafast charge transfer \cite{ChargeTransfer,EnergyTransfer,HotILE,SHG_CT,CT_Theory1,CT_Theory2}, the associated interlayer exciton formation \cite{HotILE,CT_Theory2,Merkl,PasqualILE,ILE_theory}, and energy transfer \cite{EnergyTransfer} have been extensively studied, coherent coupling has been elusive in these samples due to the rapid incoherent effects. Furthermore, the high binding energy of interlayer excitons in TMD heterostructures has raised interest in these excitons and their associated dynamics for excitonic integrated circuits \cite{Merkl,ILE_review}.
In this letter, we employ multidimensional coherent spectroscopy (MDCS) to study coherent coupling dynamics and many-body effects for electron-hole pairs, including intralayer excitons and free, unbound, interlayer electron-hole pairs in a MoSe\textsubscript{2}/WSe\textsubscript{2} heterostructure. A simple scheme of MDCS is shown in Fig.\,\ref{fig:Fig1}(a).

\begin{figure*}[t]
\includegraphics[width=0.95\textwidth]{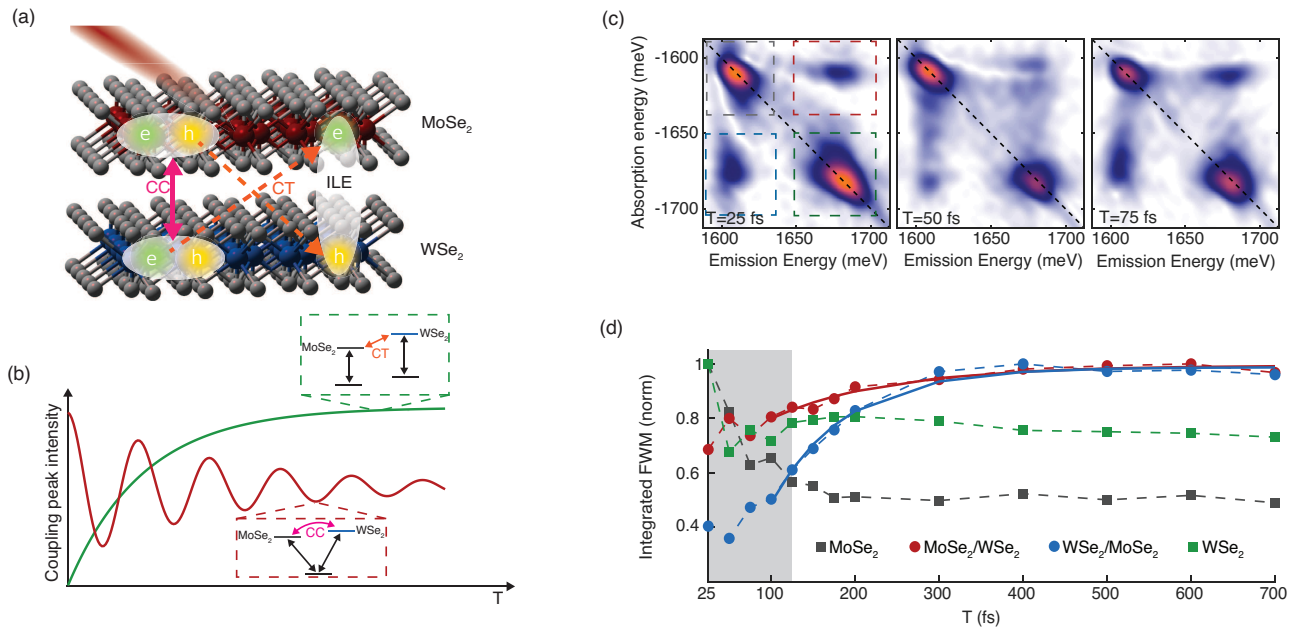}
\caption{\label{fig:Fig2} (a) Illustration of coherently coupled MoSe\textsubscript{2} and WSe\textsubscript{2} excitons and the dephasing of coherent coupling (CC) via incoherent charge transfer (CT), yielding interlayer excitons (ILE) in the heterostructure. (b) Characteristic coupling peak amplitude dynamics and associated level-systems for incoherent vs. coherent coupling. (c) Characteristic low-temperature, low-power multidimensional coherent spectra of the MoSe\textsubscript{2}/WSe\textsubscript{2} heterostructure at time-delays T=25\,fs, T=50\,fs, and T=75\,fs. All spectra are plotted on the same color scale, shown for the spectrum taken at $T$\,=\,600\,fs in Fig.\,\ref{fig:Fig1}\,(d). (d) Integrated four-wave mixing of the four peaks in the MDCS spectrum. Integration areas are marked by the dashed boxes in (c). Also shown are exponential fits to the rise behavior of the two coupling peaks from 100\,fs on (solid lines).}
\end{figure*}

With MDCS, we measure the phase-resolved evolution of an induced nonlinear response. This response is induced by three pulses (A,B,C) impinging on the sample while a fourth pulse (not shown) is used for heterodyne detection. The four pulses are separated by delays $\tau$, $T$, and $t$. Recording the phase-resolved response along $\tau$ and $t$, and Fourier transforming yields (one-quantum) spectra correlating absorption and emission energies \cite{MONSTR,Eric}, which is useful for measuring homogeneous and inhomogeneous linewidths, as well as many-body effects. Scanning the delay between the second and third pulse, $T$, reveals the temporal dynamics, similar to pump-probe, and distinguishes between coherent and incoherent coupling between different excitations \cite{MDCS_review}. 
For most of the results here we use these one-quantum spectra. For zero-quantum spectra we set $\tau=0$\,ps, move the $T$ stage and perform a subsequent Fourier transform to obtain the mixing energy axis.
Because the resonances in these heterostructure samples are very broad, even at the cryogenic temperatures and low powers we use, we add up two different measurements: A one-quantum rephasing measurement, in which the first pump interaction is conjugate, and a one-quantum non-rephasing measurement where the second pump interaction is conjugate, generating absorptive spectra \cite{MDCS_review}. Adding these two measurements together allows us to further discern different effects within this heterostructure.

We cool the sample to a temperature of 5\,K and use 100\,fs pulses at a pulse energy of 0.2\,pJ/beam and spot diameters of 1\,$\mu$m to generate the third order response. Further technical details about the experiment can be found in the Supplemental Material \cite{Supplement}.

A brightfield microscope image of the sample is provided in Fig.\,\ref{fig:Fig1}\,(b).
The sample consists of mechanically exfoliated MoSe\textsubscript{2} and WSe\textsubscript{2} monolayers stacked on top of each other with near-zero twist angle. The heterostructure is encapsulated in hexagonal boron nitride (hBN).
A photoluminescence (PL) spectrum is shown in Fig.\,\ref{fig:Fig1}\,(c). We also indicate the laser spectrum, which covers both the MoSe\textsubscript{2}- and  WSe\textsubscript{2}-A exciton resonances. 
The MDCS spectrum in Fig.\,\ref{fig:Fig1}\,(d), taken at $T=600$\,fs shows four peaks. The two peaks on the diagonal (dashed line) correspond to the resonance of the respective monolayers, while the two cross (or coupling) peaks with different absorption and emission energies are indicative of coupling between the two resonances. The spectral shift between features in PL and MDCS can be traced back to a combination of Stokes shift, spatial inhomogeneity, and limited bandwidth of the laser. From the roughly round shape of the on-diagonal peaks, we can deduce that the linewidths in this heterostructure sample are limited by the homogeneous linewidth. 
To confirm this conclusion, we fit cross-diagonal and on-diagonal slices \cite{Fits} (Supplementary Information). From these fits we are able to extract the homogeneous linewidths for the two resonances, $\gamma_{\mathrm{MoSe_2}}$=8.5\,meV and $\gamma_{\mathrm{WSe_2}}$=16.9\,meV, which are an order of magnitude larger than for monolayer samples \cite{Eric_TMD,Galan_Wse2}. It is reasonable to assume that this increase is due to additional population decay channels, such as charge transfer, in the heterostructure.

Discerning any other coupling between the excitons from the dominant charge transfer contribution is important. The key to this is the temporal evolution for the off-diagonal coupling peaks in the MDCS spectrum shown in Fig.\,\ref{fig:Fig1}\,(d).
There are two likely sources for the occurence of these peaks:

1) After excitation by a laser, electrons and holes bind together to form excitons in the MoSe\textsubscript{2} and WSe\textsubscript{2} respectively, as shown in Fig.\,\ref{fig:Fig2}\,(a). These excitons can coherently interact (coherent coupling), which can be caused by static dipole-dipole, exchange interactions, or transition dipole (F\"{o}rster) coupling \cite{CoherentCoupling} among other things. 
This coupling manifests itself through an oscillating amplitude along $T$ whose frequency matches the energy difference between the resonances as illustrated in Fig.\,\ref{fig:Fig2}(b). Decay of the oscillations stems from dephasing processes. This coupling can be thought of as a Raman-like nonradiative coherent superposition of the MoSe\textsubscript{2} and WSe\textsubscript{2} states during $T$, where the two states are coupled via a common ground state.
2) Incoherent coupling channels, such as energy or charge transfer can also lead to an appearance of the coupling peaks. Charge transfer via electron and hole transfer as illustrated in Fig.\,\ref{fig:Fig2}(a), with subsequent formation of the interlayer exciton (ILE) is common in these TMD heterostructures.
For this coupling, separate level systems do not need to share a common ground state in contrast to the coherent coupling, as illustrated in Fig.\,\ref{fig:Fig2}(b). 
Here, the coupling manifests iteself via a rise of the peak amplitude having a time scale characteristic of the transfer for these processes. 
An in-depth discussion of the different level systems can be found in the Supplemental Material \cite{Supplement}.

To better resolve the dynamics of the heterostructure and determine the coupling mechanisms, we take several MDCS spectra with varying $T$ delay. We show exemplary spectra taken at $T=$\,25\,fs, 50\,fs, and 75 fs in Fig.\,\ref{fig:Fig2}\,(c).
The most notable changes between these spectra are a visible decay for the two on-diagonal peaks and varying amplitude for the coupling peaks. Overall, the coupling peak amplitude increases in time (compare Fig.\,\ref{fig:Fig1}\,(d)). To better visualize these dynamics we integrate over each of the four peaks (integration area indicated by the dashed rectangles in Fig.\,\ref{fig:Fig2}\,(c)), and plot the resulting integrated amplitudes in Fig.\,\ref{fig:Fig2}\,(d).
The decay of the on-diagonal peaks (squares) as well as the rise of the coupling peaks (circles) is clearly visible. Moreover, the amplitude of the two cross-peaks is non-zero at early times and shows features suggestive of oscillations, both indications of coherent coupling.
The decay for both on-diagonal peaks occurs rapidly and noticeably faster than the rise of the coupling peaks. This decay can be explained by the multitude of processes, including rapid decay into dark or localized states as reported for both materials in the literature \cite{Galan_Wse2,Kasprzak} and population decay into the ground state, all of which affect the on-diagonal peaks.
The rise of the coupling peaks, however, occurs only  due to processes that incoherently couple the two materials together, such as energy and charge transfer. Based on the extensive literature on these heterostructures \cite{ChargeTransfer,ChargeTranfer2,EnergyTransfer,SHG_CT} and the similar time-scale of the rise for the two coupling peaks, we deduce charge transfer to be the dominant incoherent coupling mechanism in this sample. 

We fit an inverse exponential for $T\geq$100\,fs to the rise of the coupling peaks (solid lines) which yields an estimated rise time $\tau_{M\rightarrow W}<$ 149$\pm$28\,fs for the MoSe\textsubscript{2}/WSe\textsubscript{2} peak and $\tau_{W\rightarrow M}<$ 91$\pm$9\,fs for the WSe\textsubscript{2}/MoSe\textsubscript{2} peak, limited by the temporal resolution of 100\,fs in our experiment (Supplementary Information). These values are in good agreement with the charge transfer times in the literature for similar samples \cite{ChargeTransfer,ChargeTranfer2,EnergyTransfer,SHG_CT}. 

\begin{figure}
\includegraphics[width=0.5\textwidth]{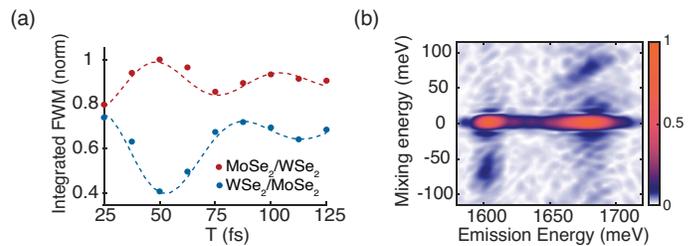}
\caption{\label{fig:Fig3} (a) High temporal resolution measurement of the two-coupling peaks from Fig.\,\ref{fig:Fig2}(b). Curves have been offset to increase readability. The MoSe\textsubscript{2}/WSe\textsubscript{2} curve was recorded with positive intentional GDD, the WSe\textsubscript{2}/MoSe\textsubscript{2} curve was recorded with negative intentional GDD. (b) Zero-quantum spectrum, similar to Fourier-transforming with respect to the delay of a spectrally-resolved pump-probe experiment. Here, we added measurements taken with positive and negative intentional GDD. Complete data sets can be found in the Supplemental Material \cite{Supplement}.}
\end{figure}

A separate data set with smaller $T$ steps (range indicated by the gray area in Fig.\,\ref{fig:Fig2}\,(d)), shown in Fig.\,\ref{fig:Fig3}\,(a), resolves early time dynamics better. Both coupling peaks show clear oscillations with a frequency around $\hbar \omega=$\,74\,meV, corresponding to the energy difference between the MoSe\textsubscript{2} and WSe\textsubscript{2} excitons. These oscillations are emphasized by plotting a decaying cosine together with an exponential rise (dashed lines) as a guide to the eye. Since we see signatures of both coherent coupling and incoherent charge transfer, the resulting plots in Fig.\,\ref{fig:Fig2}(d) and Fig.\,\ref{fig:Fig3}(a) are the sum of a decaying oscillation and an exponential rise, as expected from Fig.\,\ref{fig:Fig2}(b).

A more quantitative measure of these oscillations can be obtained by taking a zero-quantum MDCS spectrum \cite{ZeroQuantum}. It is similar to spectrally-resolved pump-probe spectroscopy with an additional Fourier transform along the pump-probe delay $T$, which yields the mixing energy axis. Because we have used phase-resolved heterodyne detection, we can resolve the sign of the oscillation, which is different for the two coupling features in the zero-quantum spectrum shown in Fig. \ref{fig:Fig3}\,(b).
The two features at zero mixing energy and MoSe\textsubscript{2} or WSe\textsubscript{2} emission energy correspond to the non-oscillating contributions of both on-diagonal and coupling peaks. Two features around a mixing energy of +71\,meV and -74\,meV can be seen for the MoSe\textsubscript{2} and WSe\textsubscript{2} emission, respectively. Within the sample inhomogeneity, this matches the energy spacing between the MoSe\textsubscript{2} and WSe\textsubscript{2}-A excitons. This agreement supports the fact that excitons in the two materials are indeed coherently coupled and oscillate during $T$ with a frequency determined by the energy difference of the two resonances. The broadness of the features stems from the fact that both coupling contributions decay rapidly along $T$. The low energy (10-20\,meV) signatures are due to truncation effects.
The nature of the coherent coupling is not immediately evident and needs further investigation. Common interactions that lead to coherent coupling are biexcitonic of nature, such as static dipole-dipole or exchange interactions, or due to mixing of the single exciton states, such as transition dipole (F\"{o}rster) coupling \cite{ZhangPNAS,CoherentCoupling}. They can, for example, be distinguished by careful analysis of the real part of the MDCS spectrum, as demonstrated in \cite{CoherentCoupling}. Future experiments using double-quantum coherent spectroscopy might also provide insight into the nature of the coherent coupling \cite{QW1,QW2}.
In Fig.\,\ref{fig:Fig3}\,(a) and (b), we combine measurements with intentional small negative and positive group-delay dispersion (GDD) on the three excitation pulses, which enhances the coherent oscillations that are otherwise obscured by residual third order dispersion. A detailed discussion using simulations of the optical Bloch equations can be found in the Supplementary Information.

\begin{figure}
\includegraphics[width=0.5\textwidth]{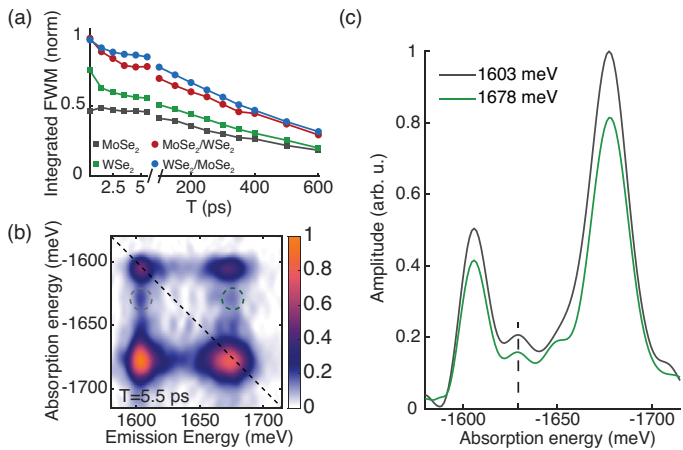}
\caption{\label{fig:Fig4} (a) Integrated four-wave mixing of the four peaks in the MDCS spectrum for intermediate and long times (500\,fs-600\,ps). (b) Characteristic low-temperature, low-power multidimensional coherent spectrum of the MoSe\textsubscript{2}/WSe\textsubscript{2} heterostructure at a pump-probe time-delay of T=5.5\,ps. (c) Slices of fixed emission energy taken along the absorbtion energy axis. The dashed line indicates the absorption energy for the interlayer exciton feature.}
\end{figure}

We furthermore acquired MDCS spectra with varying $T$ delay over an intermediate (500\,fs-5\,ps) to longterm (150\,ps-600\,ps) time range, whose integrated peak amplitudes are shown in Fig.\,\ref{fig:Fig4}\,(a), starting at 500\,fs, after the initial coupling peak rise and on-diagonal peak decay. The full dataset can be found in the Supplementary Information.

Within the first 5\,ps, three of the four peaks (except the MoSe\textsubscript{2} peak) decay, followed by another decay for all four peaks for $T\geq$100\,ps. This behavior is consistent with the literature, which reports bi-exponential time-scales in the intermediate and long term dynamics of these samples \cite{HotILE,CT_Theory2,SHG_CT}.
In contrast to previous experiments however, which look at the coupling dynamics in an isolated setting, the MDCS approach reveals distinctly different decay constants for the four peaks, hinting at superposition of dynamics that cannot be obtained in the selective experiments in the literature. In the intermediate time frame, the MoSe\textsubscript{2}/WSe\textsubscript{2} peak ($\tau$=1.22$\pm$0.12\,ps), as well as the WSe\textsubscript{2} peak ($\tau$=1.51$\pm$0.12\,ps) decay quicker and with a larger amplitude than the WSe\textsubscript{2}/MoSe\textsubscript{2} peak ($\tau$=1.69$\pm$0.16\,ps) and MoSe\textsubscript{2} peak (no fit). Here, $\tau$ denotes the decay constant from a simple exponential decay.
Competition between several dynamical processes can explain this surprising behavior.
We attribute the initial decay of the coupling peaks to the relaxation of hot interlayer excitons, which usually live for 600-800\,fs before decaying into a cold, bound interlayer exciton state \cite{HotILE,CT_Theory2}. Either the difference in many-body effects of the hot vs. cold interlayer excitons, or the fact that the hot interlayer excitons are more prone to electron-hole recombination than their tightly bound counterparts can explain an initial decay of the coupling peaks. 
The lack of decay, and even slight rise, of the MoSe\textsubscript{2} peak could be explained by an exciton population increase due to coupling back from dark states. Additional processes such as energy transfer from WSe\textsubscript{2} into the MoSe\textsubscript{2} can also contribute on these intermediate time-scales.

The long term decay, which is fitted separately due to competing dynamics at early times, shows a roughly uniform timescale for all four peaks ($607\pm22$\,ps, $580\pm28$\,ps, $540\pm8$\,ps ,$520\pm13$\,ps from left to right, top to bottom) within the uncertainty of the measurement and noise. We attribute this decay to interlayer exciton decay. The decay time is in good agreement with the literature, which has reported interlayer exciton lifetimes from hundreds of picoseconds, to hundreds of nanoseconds, depending on the twist angle and other sample parameters \cite{PasqualILE,GalanILE,ILE_lifetime1,ILE_lifetime2,ILE_lifetime3}.

We have reproduced all the dynamics in this sample, including charge transfer, coherent coupling, and hot interlayer exciton relaxation in a similar MoSe\textsubscript{2}/WSe\textsubscript{2} heterostructure (Supplementary Information), showing that the occurence of these dynamics is not sample specific. 

After the relaxation of the hot interlayer excitons, another signature of the interlayer exciton is obtained in the MDCS spectra. An MDCS spectrum with a $T$ delay of 5.5\,ps is shown in Fig.\,\ref{fig:Fig4}\,(b). The features that show absorption at continuum energies around 1629\,meV and emission at the MoSe\textsubscript{2} and WSe\textsubscript{2}-A exciton resonances (dashed circles), have no visible corresponding on-diagonal peaks, and no emission at comparable energies. Therefore these spectral features cannot be due to another material resonance. Instead, we believe that this features occurs due to the continuum absorption generating free interlayer electron/hole pairs, which subsequently affect the emission of the MoSe\textsubscript{2} and WSe\textsubscript{2} excitons via many-body effects. The weak absorption by the free interlayer electron-hole pairs is compensated by their very strong Coulomb interaction with the excitons. Similar effects have been observed in GaAs quantum wells \cite{ZhangPNAS,Borca_MBE}. 
Two slices, taken at fixed emission energies of 1603\,meV and 1678\,meV along the absorption energy axis are displayed in Fig.\,\ref{fig:Fig4}\,(c). From this, we determine the interlayer exciton absorption feature to be at 1629\,meV. From our PL measurement, the ILE emission is known to be at 1.375\,eV (Supplementary Information). 
From the difference between these energies we deduce the binding energy of the interlayer excitons in this sample to be around 254\,meV. This is in excellent agreement with a recently performed first principle calculation \cite{ILE_theory}, which estimates the binding energy to be around 250\,meV. This binding energy is consistent with previously measured binding energies in these samples at 110\,K using microARPES and PL \cite{BindingXD}, and two times larger than measured binding energies in a WSe\textsubscript{2}/WS\textsubscript{2} heterostructure \cite{Merkl}.

In summary, we demonstrated coherent exciton-exciton coupling, rapid charge transfer, and tightly bound interlayer excitons with binding energies above 200\,meV on a mixed TMD heterostructure. These results underscore the immense technological potential for these materials. Studying the nature of the coherent coupling as well as how to tune it will remain an exciting challenge for future work. 

\medskip
\begin{acknowledgments}
We thank M.W. Day for fruitful discussion. The research at U. of Michigan was supported by NSF Grant No. 2016356. The work at U. Washington was supported by the Department of Energy, Basic Energy Sciences, Materials Sciences and Engineering Division (DE-SC0018171). W.G.H. was supported by the NSF Graduate Research Fellowship Program under Grant No. DGE-1762114.
\end{acknowledgments}

\bibliography{Library}

\clearpage

\end{document}